\documentclass[aps,prb,preprint,amsmath,amssymb]{revtex4-2}

\usepackage{graphicx}
\usepackage{bm}
\usepackage{hyperref}

\begin{document}

\title{Tangent-Plane Evidential Uncertainty in Active Learning for Magnetic Interatomic Potentials}

\author{Yang Cheng}
\affiliation{Key Laboratory of Computational Physical Sciences (Ministry of Education), Institute of Computational Physical Sciences, State Key Laboratory of Surface Physics, and Department of Physics, Fudan University, Shanghai, 200433, China}

\author{Hongyu Yu}
\email[Corresponding author: ]{hongyuyu20@fudan.edu.cn}
\affiliation{Key Laboratory of Computational Physical Sciences (Ministry of Education), Institute of Computational Physical Sciences, State Key Laboratory of Surface Physics, and Department of Physics, Fudan University, Shanghai, 200433, China}

\author{Hongjun Xiang}
\email[Corresponding author: ]{hxiang@fudan.edu.cn}
\affiliation{Key Laboratory of Computational Physical Sciences (Ministry of Education), Institute of Computational Physical Sciences, State Key Laboratory of Surface Physics, and Department of Physics, Fudan University, Shanghai, 200433, China}

\begin{abstract}
Magnetic interatomic potentials need to account for coupled lattice and spin degrees of freedom, yet constructing reliable training sets remains costly because noncollinear first-principles labels are expensive. Active learning can mitigate this cost, provided that the uncertainty estimate is physically meaningful for the magnetic-response targets that drive spin reorientation. Here we introduce Equivariant Evidential Deep Learning for Spin-Lattice Potentials ($\mathrm{e}^2\mathrm{SLP}$), a magnetic uncertainty-learning method that formulates the projected spin-force likelihood and the corresponding epistemic uncertainty in the tangent plane orthogonal to the local spin direction. This geometry is consistent with the constrained-moment labels, which contain no independent radial response, and yields a structure-level projected-spin-force uncertainty score for candidate selection. Using bulk BiFeO$_3$ and monolayer CrTe$_2$ as benchmark systems, we show that this score correlates strongly with prediction errors and that the full $\mathrm{e}^2\mathrm{SLP}$ selection, training, and fine-tuning workflow lowers energy, force, and projected spin-force errors relative to random sampling. For monolayer CrTe$_2$, round-by-round comparisons show accuracy competitive with a four-model committee, while a four-seed ablation separates the contributions of uncertainty-guided selection and $\mathrm{e}^2\mathrm{SLP}$ training followed by fine-tuning. $\mathrm{e}^2\mathrm{SLP}$ thus provides a single-model tangent-plane uncertainty criterion for data-efficient construction of magnetic machine-learning interatomic potentials.
\end{abstract}

\maketitle

\section{Introduction}
Magnetic materials are central to information storage, spintronics, magneto-optical technologies, and finite-temperature materials modeling\cite{vedmedenko_2020_2020,hirohata_review_2020}. In many systems, especially those with strong coupling among lattice, spin, and magnetic anisotropy, reliable simulation requires explicit magnetic degrees of freedom rather than an effective nonmagnetic approximation\cite{dudarev_magnetic_2005,ma_large-scale_2008,ma_spin-lattice-electron_2012,tranchida_massively_2018,yang_deep_2024,kotykhov_actively_2025,rinaldi_non-collinear_2024,yu_spin-dependent_2024,yu_physics-informed_2024,zhang_fully_2026}. Machine-learning interatomic potentials (MLIPs) extend first-principles accuracy to larger length and time scales\cite{behler_generalized_2007,bartok_gaussian_2010,shapeev_moment_2016,batzner_e3-equivariant_2022,batatia_mace_2022,deng_chgnet_2023}, and recent magnetic MLIPs, including SpinGNN\cite{yu_spin-dependent_2024}, mMTP, and mMACE, have begun to incorporate local magnetic moments or spin degrees of freedom directly into the model\cite{novikov_magnetic_2022,chapman_machine-learned_2022,yang_deep_2024,yu_physics-informed_2024,kotykhov_fitting_2024,xu_spin-informed_2025,ho_equivariant_2026}. Their construction is particularly demanding because the training data need to cover both structural and spin configurational spaces, while reference labels often require expensive constrained noncollinear density-functional-theory calculations\cite{ma_constrained_2015,kotykhov_constrained_2023,zheng_integrating_2026}.

This data-construction bottleneck has motivated active-learning strategies for interatomic potentials\cite{chen_data-driven_2026,podryabinkin_active_2017,PhysRevMaterials.3.023804,zhang_dp-gen_2020,lin_searching_2021,jung_active_2024,bidoggia_automated_2025,henkes_aims-pax_2025}. Existing workflows use extrapolation grades, committee or variance estimates, descriptor-space diversity, or hybrid acquisition rules to reduce manual dataset curation and identify informative configurations\cite{chen_data-driven_2026,podryabinkin_active_2017,lin_searching_2021,jung_active_2024,bidoggia_automated_2025,henkes_aims-pax_2025}. Active learning has also been applied to training-set construction for magnetic interatomic potentials\cite{kotykhov_actively_2025}. Practical indicators such as extrapolation grades can be effective\cite{podryabinkin_active_2017}, but they are not formulated as uncertainty models for the projected magnetic-response targets considered here. Committee-based criteria require training multiple models and therefore add substantial cost to iterative SOC-aware active-learning loops\cite{PhysRevMaterials.3.023804,zhang_dp-gen_2020,schran_committee_2020,wollschlager_uncertainty_2023}.

A closely related development is Equivariant Evidential Deep Learning for Interatomic Potentials ($\mathrm{e}^2\mathrm{IP}$), a single-model uncertainty framework for vector-valued interatomic-potential targets such as atomic forces\cite{wang_equivariant_2026}. In this framework, the model predicts not only the force vector but also a full rotation-consistent $3\times3$ symmetric positive-definite covariance tensor, and uses a Normal--Inverse--Wishart evidential prior to separate epistemic and aleatoric uncertainty\cite{wang_equivariant_2026,amini_deep_nodate,meinert_multivariate_2022,xu_evidential_2025}. This makes $\mathrm{e}^2\mathrm{IP}$ attractive for ranking candidate configurations in active learning, because uncertainty can be obtained from a single forward pass rather than from an ensemble. However, directly applying this framework to magnetic spin-force learning would be geometrically inconsistent. In the present magnetic setting, the learned spin-response target is not a fully unconstrained three-dimensional vector but the projected spin-force target $\mathbf f_{s,\perp}$ in the tangent plane orthogonal to the local spin direction\cite{yu_physics-informed_2024,kotykhov_fitting_2024}. Because the constrained-moment labels contain no independent radial response, a three-dimensional uncertainty model would introduce an uncertainty component without a corresponding physical target. Using the evidential idea for magnetic interatomic potentials therefore requires a new uncertainty geometry that matches the tangent-space structure of spin-force prediction.

In this work, we introduce Equivariant Evidential Deep Learning for Spin-Lattice Potentials ($\mathrm{e}^2\mathrm{SLP}$), a SOC-aware magnetic interatomic-potential method inspired by the evidential construction of $\mathrm{e}^2\mathrm{IP}$, in which the likelihood, predictive moments, and epistemic uncertainty of projected spin forces are formulated in the two-dimensional tangent plane\cite{wang_equivariant_2026}. This formulation matches the transverse response supplied by constrained-moment DFT. The resulting magnetic-site-averaged epistemic trace defines a structure-level projected-spin-force uncertainty score, $U_{\mathrm{epi}}^{\mathrm{sf}}$, for candidate selection. We evaluate this score and the associated selection, training, and fine-tuning workflow using bulk BiFeO$_3$ and monolayer CrTe$_2$ as benchmark systems. Across both systems, $U_{\mathrm{epi}}^{\mathrm{sf}}$ correlates strongly with prediction error and the full $\mathrm{e}^2\mathrm{SLP}$ workflow lowers energy, force, and projected spin-force errors relative to random sampling. For CrTe$_2$, we also compare with a four-model committee baseline across active-learning rounds and use a four-seed ablation on the same system to distinguish uncertainty-guided selection from the $\mathrm{e}^2\mathrm{SLP}$ training and fine-tuning route.

\section{Method}
\begin{figure*}
    \centering
    \includegraphics[width=\textwidth]{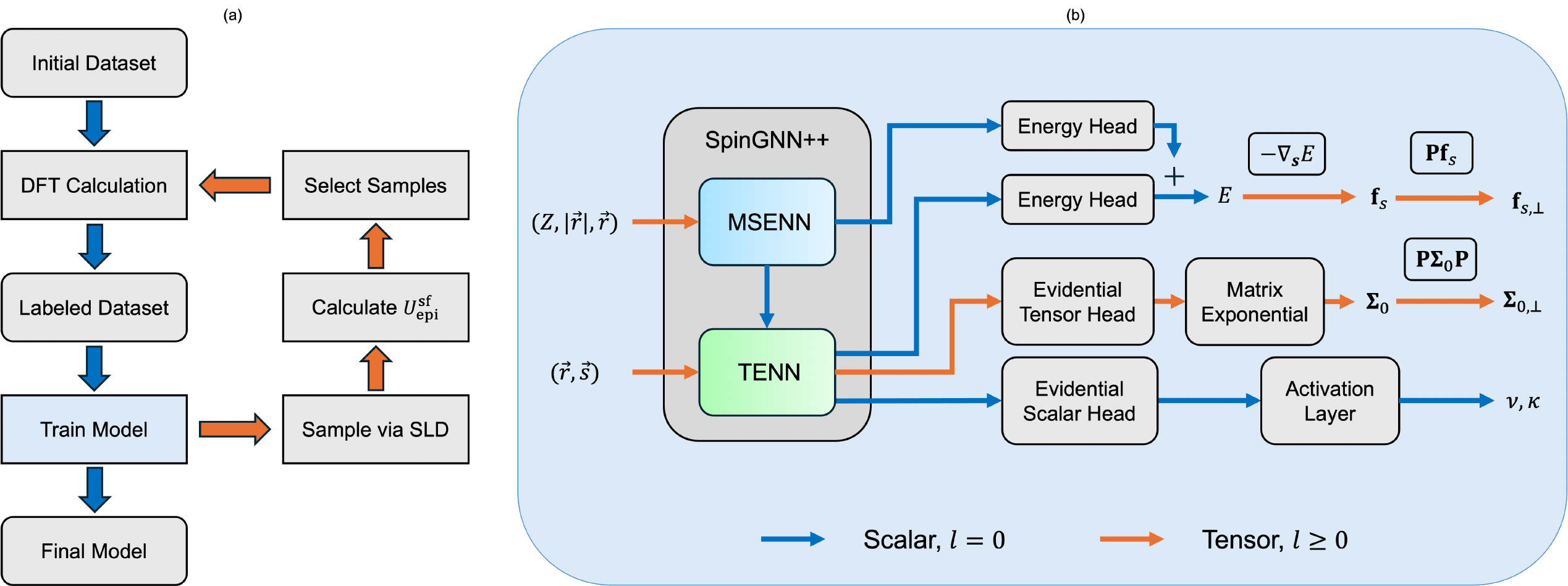}
    \caption{$\mathrm{e}^2\mathrm{SLP}$ active-learning framework. (a) Uncertainty-guided active-learning workflow: starting from an initial dataset, reference DFT calculations produce a labeled dataset for model training; the trained model is then used together with candidate magnetic configurations generated by spin-lattice dynamics (SLD) to evaluate the structure-level uncertainty score $U_{\mathrm{epi}}^{\mathrm{sf}}$, select high-uncertainty configurations for additional labeling, and retrain the model on the expanded dataset. (b) $\mathrm{e}^2\mathrm{SLP}$ architecture: the SpinGNN++-inspired backbone, comprising the magnetic-symmetry-preserving (MSENN) and time-reversal-equivariant (TENN) channels, combines scalar and tensor equivariant channels to construct the total energy, from which the raw spin force is obtained and projected onto the tangent plane; the evidential tensor and scalar branches provide the canonical uncertainty-shape matrix and evidential strength parameters used to form the tangent-plane-consistent projected-spin-force epistemic uncertainty. Blue and orange arrows denote scalar ($l=0$) and tensor ($l\ge 0$) channels, respectively.}
    \label{fig:framework}
\end{figure*}

Our method combines an uncertainty-guided active-learning workflow with the $\mathrm{e}^2\mathrm{SLP}$ model. Starting from an initial constrained-DFT-labeled dataset, the evidential model evaluates candidate magnetic configurations generated by spin-lattice dynamics and selects configurations with large projected-spin-force uncertainty for additional first-principles labeling. The $\mathrm{e}^2\mathrm{SLP}$ model uses an SOC-aware magnetic energy backbone and a tangent-plane evidential uncertainty branch: atomic forces and projected spin forces are derived from a shared spin--lattice energy surface, while the projected-spin-force epistemic covariance is evaluated in the local tangent plane. Its magnetic-site-averaged trace defines the structure-level selection score $U_{\mathrm{epi}}^{\mathrm{sf}}$. The first-principles labels were obtained from DFT calculations with spin--orbit coupling and DFT+$U$ corrections\cite{kresse_efficiency_1996,blochl_projector_1994,perdew_generalized_1996,dudarev_electron-energy-loss_1998,liechtenstein_density-functional_1995}; system-specific computational settings are given in the Supplemental Material\cite{supplemental_material}. The overall workflow and model architecture are summarized in Fig.~\ref{fig:framework}.

\subsection{SpinGNN++-inspired magnetic backbone}
Our magnetic interatomic potential follows the SpinGNN++ framework\cite{yu_physics-informed_2024}. MSENN describes explicit spin--lattice interactions, while TENN learns higher-order interactions with time-reversal equivariance. The total energy is
\begin{equation}
E^{\mathrm{total}}(\mathbf r,\mathbf s)=E^{\mathrm{MSENN}}(\mathbf r,\mathbf s)+E^{\mathrm{TENN}}(\mathbf r,\mathbf s),
\label{eq:spingnn-total-energy}
\end{equation}
where $\mathbf r$ denotes atomic coordinates and $\mathbf s$ denotes local spin variables.

For the present work, we adopt this SpinGNN++ perspective as the architectural foundation and implement the model through an SOC-aware Spin-Allegro-style local-environment network\cite{musaelian_learning_2023}. The learned total energy can be written as
\begin{equation}
E_{\mathrm{tot}} = E_{\mathrm{struct}} + E_J + E_{\mathrm{ani}} + E_{\mathrm{biquad}} + E_{\mathrm{SOC}},
\label{eq:model-energy-decomposition}
\end{equation}
where the separate terms represent the structural interaction energy, exchange-related magnetic coupling, anisotropy contribution, biquadratic magnetic term, and SOC-related contribution, respectively. The model is trained on a single spin--lattice energy surface, from which all response observables are derived.

\subsection{Energy-derived forces and projected spin forces}
Atomic forces are defined as the negative energy gradient with respect to atomic coordinates, whereas spin-force-related quantities are defined from the energy derivative with respect to the local spin-direction unit vector. Specifically, the spin coordinate used in the derivative is the dimensionless direction $\hat{\mathbf s}_i$ of the local magnetic moment, not the magnetic-moment magnitude. The raw spin-force quantity is therefore $\mathbf f_{s,i}=-\partial E/\partial\hat{\mathbf s}_i$ and has units of energy. Denoting the raw spin-force quantity by $\mathbf f_s$, the physically relevant projected spin force is
\begin{equation}
\mathbf f_{s,\perp}=\mathbf P\mathbf f_s, \qquad \mathbf P = \mathbf I - \hat{\mathbf s}\hat{\mathbf s}^{\top},
\label{eq:projected-spin-force}
\end{equation}
where $\hat{\mathbf s}=\mathbf s/\|\mathbf s\|$ is the normalized local spin direction. This projection removes the radial component parallel to the spin direction and retains only the tangent component compatible with a unit-vector spin coordinate.

The magnetic-supervision labels were obtained from constrained noncollinear DFT calculations with spin--orbit coupling, as detailed in the Supplemental Material\cite{supplemental_material}. The resulting constraining field is transverse to the prescribed local spin direction\cite{ma_constrained_2015}. We use this signal as the projected spin-force supervision target $\mathbf f_{s,\perp,i}^{\mathrm{DFT}}$, with the convention given in the Supplemental Material\cite{supplemental_material}, and formulate the uncertainty model for the same projected quantity. We therefore treat the tangent-plane formulation as a consistency requirement imposed by the constrained-spin target, rather than as an empirical model choice among three-dimensional alternatives.

\subsection[e2SLP formulation for magnetic spin-force uncertainty]{$\mathrm{e}^2\mathrm{SLP}$ formulation for magnetic spin-force uncertainty}
To quantify predictive uncertainty, $\mathrm{e}^2\mathrm{SLP}$ adapts the equivariant evidential construction introduced in $\mathrm{e}^2\mathrm{IP}$ to the magnetic spin-force setting. Following Wang \emph{et al.}\cite{wang_equivariant_2026} and the broader evidential-regression perspective of Amini \emph{et al.}\cite{amini_deep_nodate}, each target vector $\mathbf y_i$ is described through a Gaussian likelihood with latent mean and covariance,
\begin{equation}
p(\mathbf y_i \mid \boldsymbol\mu_i, \boldsymbol\Sigma_i)=\mathcal N(\mathbf y_i \mid \boldsymbol\mu_i, \boldsymbol\Sigma_i),
\label{eq:gaussian-likelihood}
\end{equation}
while the pair $(\boldsymbol\mu_i,\boldsymbol\Sigma_i)$ is assigned a Normal--Inverse--Wishart prior,
\begin{equation}
\boldsymbol\Sigma_i \sim \mathcal W^{-1}(\boldsymbol\Psi_i,\nu_i), \qquad
\boldsymbol\mu_i \mid \boldsymbol\Sigma_i \sim \mathcal N\!\left(\boldsymbol\gamma_i,\frac{1}{\kappa_i}\boldsymbol\Sigma_i\right).
\label{eq:normal-inverse-wishart-prior}
\end{equation}
We adopt the scale parameterization used in $\mathrm{e}^2\mathrm{IP}$,
\begin{equation}
\boldsymbol\Psi_i \triangleq \nu_i \boldsymbol\Sigma_{0,i},
\label{eq:e2ip-scale-parameterization}
\end{equation}
where $\boldsymbol\Sigma_{0,i}$ is an SPD matrix that acts as the canonical uncertainty-shape matrix.

After marginalizing the latent variables, the predictive distribution becomes multivariate Student-$t$, and the uncertainty decomposes into aleatoric and epistemic parts,
\begin{equation}
\mathbf U_{\mathrm{ale}} = \mathbb E[\boldsymbol\Sigma_i] = \frac{\nu_i \boldsymbol\Sigma_{0,i}}{\nu_i-d-1},
\label{eq:aleatoric-uncertainty}
\end{equation}
\begin{equation}
\mathbf U_{\mathrm{epi}} = \mathrm{Var}[\boldsymbol\mu_i] = \frac{1}{\kappa_i}\mathbb E[\boldsymbol\Sigma_i] = \frac{\nu_i \boldsymbol\Sigma_{0,i}}{\kappa_i(\nu_i-d-1)}.
\label{eq:epistemic-uncertainty}
\end{equation}
We retain the full-covariance $\mathrm{e}^2\mathrm{IP}$ formulation but evaluate it for the projected spin-force target in the corresponding tangent plane.

As in the original $\mathrm{e}^2\mathrm{IP}$ framework, the covariance tensor is constructed from an equivariant symmetric representation and mapped to a strictly SPD matrix through the matrix exponential,
\begin{equation}
\boldsymbol\Sigma_{0,i}=\exp(\mathbf S_i).
\label{eq:canonical-covariance}
\end{equation}
We then project this canonical uncertainty-shape matrix into the tangent subspace,
\begin{equation}
\boldsymbol\Sigma_{0,i,\perp}=\mathbf P_i\boldsymbol\Sigma_{0,i}\mathbf P_i.
\label{eq:tangent-projected-covariance}
\end{equation}
To match the projected spin-force target, we evaluate the covariance in the local two-dimensional tangent basis, $\boldsymbol\Sigma^{2d}_{0,i}=\mathbf E_i^{\top}\boldsymbol\Sigma_{0,i,\perp}\mathbf E_i$, so that the predicted uncertainty corresponds only to the transverse response supplied by constrained-moment DFT.

\subsection{Tangent-plane-consistent uncertainty geometry and training objective}
Let $\hat{\mathbf s}$ denote the normalized local spin direction. The projector $\mathbf P$ removes the component parallel to the spin axis. A three-dimensional covariance or uncertainty tensor $\boldsymbol\Sigma$ is then mapped to the tangent space through
\begin{equation}
\boldsymbol\Sigma_{\perp} = \mathbf P\boldsymbol\Sigma\mathbf P.
\label{eq:tangent-covariance-projection}
\end{equation}
To evaluate the likelihood in minimal coordinates, we introduce a local orthonormal tangent basis
\begin{equation}
\mathbf E_i=[\mathbf e_{i,1},\mathbf e_{i,2}] \in \mathbb R^{3\times 2},
\label{eq:tangent-basis}
\end{equation}
with $\mathbf e_{i,1}$ and $\mathbf e_{i,2}$ orthonormal to $\hat{\mathbf s}_i$ and to each other. In practice, $\mathbf E_i$ is constructed by Gram--Schmidt orthonormalization against a fixed Cartesian reference axis chosen non-parallel to $\hat{\mathbf s}_i$; the resulting negative log-likelihood and the trace-based uncertainty score are invariant under in-plane rotations of this basis. The reference target, predictive mean, and canonical uncertainty-shape matrix are then projected as
\begin{equation}
\mathbf y_{i,2d}=\mathbf E_i^{\top}\mathbf y_i, \qquad \boldsymbol\gamma_{i,2d}=\mathbf E_i^{\top}\boldsymbol\gamma_i, \qquad \boldsymbol\Sigma^{2d}_{0,i}=\mathbf E_i^{\top}\boldsymbol\Sigma_{0,i,\perp}\mathbf E_i.
\label{eq:tangent-coordinate-projection}
\end{equation}
The predictive mean $\boldsymbol\gamma_i$ is not an independent vector head: it is identified with the projected energy gradient with respect to the local spin-direction unit vector, $\boldsymbol\gamma_i = -\mathbf P_i\,\partial E/\partial \hat{\mathbf s}_i$. The evidential branch consequently only parameterizes the uncertainty tensor and evidence parameters, so that all response observables remain tied to a single spin--lattice energy surface.

For the projected spin-force branch we use $d=2$ rather than $d=3$, and enforce
\begin{equation}
\nu_i = \mathrm{softplus}(\hat{\nu}_i) + (d+2), \qquad \kappa_i = \mathrm{softplus}(\hat{\kappa}_i) + \epsilon,
\label{eq:evidence-parameters}
\end{equation}
so that the corresponding uncertainty moments remain valid. The training loss follows the $\mathrm{e}^2\mathrm{IP}$ multivariate Student-$t$ negative log-likelihood, evaluated on the tangent-plane coordinates,
\begin{equation}
\begin{aligned}
\mathcal L_{\mathrm{NLL},i} = &\; \log \Gamma\!\left(\frac{\nu_i-d+1}{2}\right)-\log \Gamma\!\left(\frac{\nu_i+1}{2}\right)
+\frac{d}{2}\log\!\left(\frac{\pi\nu_i(1+\kappa_i)}{\kappa_i}\right) \\
&+\frac{1}{2}\log |\boldsymbol\Sigma^{2d}_{0,i}|
+\frac{\nu_i+1}{2}\log\!\left(1+\frac{\kappa_i}{\nu_i(1+\kappa_i)}
(\mathbf y_{i,2d}-\boldsymbol\gamma_{i,2d})^{\top}(\boldsymbol\Sigma^{2d}_{0,i})^{-1}(\mathbf y_{i,2d}-\boldsymbol\gamma_{i,2d})\right),
\end{aligned}
\label{eq:tangent-nll}
\end{equation}
with an additional error-weighted evidence regularizer, specified in the Supplemental Material\cite{supplemental_material}, to discourage overconfident predictions on poorly fitted samples.

\subsection{Structure-level uncertainty score for active learning}
The tangent-plane-consistent epistemic uncertainty of the spin-force branch is used as the basis for active-learning acquisition. At the node level, we work directly with the two-dimensional epistemic covariance
\begin{equation}
\mathbf U_{\mathrm{epi},i}^{\mathrm{sf},2d}=\frac{\nu_i\boldsymbol\Sigma^{2d}_{0,i}}{\kappa_i(\nu_i-d-1)},
\label{eq:node-epistemic-spin-force}
\end{equation}
and define the structure-level scalar proxy for spin-force epistemic uncertainty for a configuration $\mathcal C$ as the magnetic-site-averaged trace,
\begin{equation}
U_{\mathrm{epi}}^{\mathrm{sf}}(\mathcal C)=\frac{1}{|\mathcal C_{\mathrm{mag}}|}\sum_{i\in\mathcal C_{\mathrm{mag}}}\mathrm{tr}\!\left(\mathbf U_{\mathrm{epi},i}^{\mathrm{sf},2d}\right).
\label{eq:structure-epistemic-score}
\end{equation}
Here $\mathcal C_{\mathrm{mag}}$ denotes the magnetic sites, Fe in BiFeO$_3$ and Cr in CrTe$_2$. Nonmagnetic atoms are excluded from this projected-spin-force uncertainty average because no local spin direction, and therefore no tangent plane, is defined for them. During iterative dataset expansion, configurations with larger $U_{\mathrm{epi}}^{\mathrm{sf}}$ are prioritized for first-principles labeling from the spin-lattice-dynamics candidate pool\cite{ma_large-scale_2008,ma_langevin_2011,tranchida_massively_2018,thompson_lammps_2022,garcia-palacios_langevin-dynamics_1998}.

For the uncertainty-guided active-learning branch, the evidential $\mathrm{e}^2\mathrm{SLP}$ model is used for uncertainty estimation and configuration selection. Final accuracy is evaluated with the same conventional magnetic-potential architecture used for the random baseline. In the $\mathrm{e}^2\mathrm{SLP}$ training/fine-tuning route, this conventional model is initialized from the shared $\mathrm{e}^2\mathrm{SLP}$ parameters and then fine-tuned on the selected dataset. This protocol is used at the final active-learning round for bulk BiFeO$_3$ and at every active-learning round for monolayer CrTe$_2$. The CrTe$_2$ ablation further compares this route with conventional training from random selection and with the same $\mathrm{e}^2\mathrm{SLP}$ training/fine-tuning route applied after random selection.

\section{Results and Discussion}
\subsection[Correlation between Uepi sf and prediction error]{Correlation between $U_{\mathrm{epi}}^{\mathrm{sf}}$ and prediction error}
\begin{figure*}
    \centering
    \includegraphics[width=\textwidth]{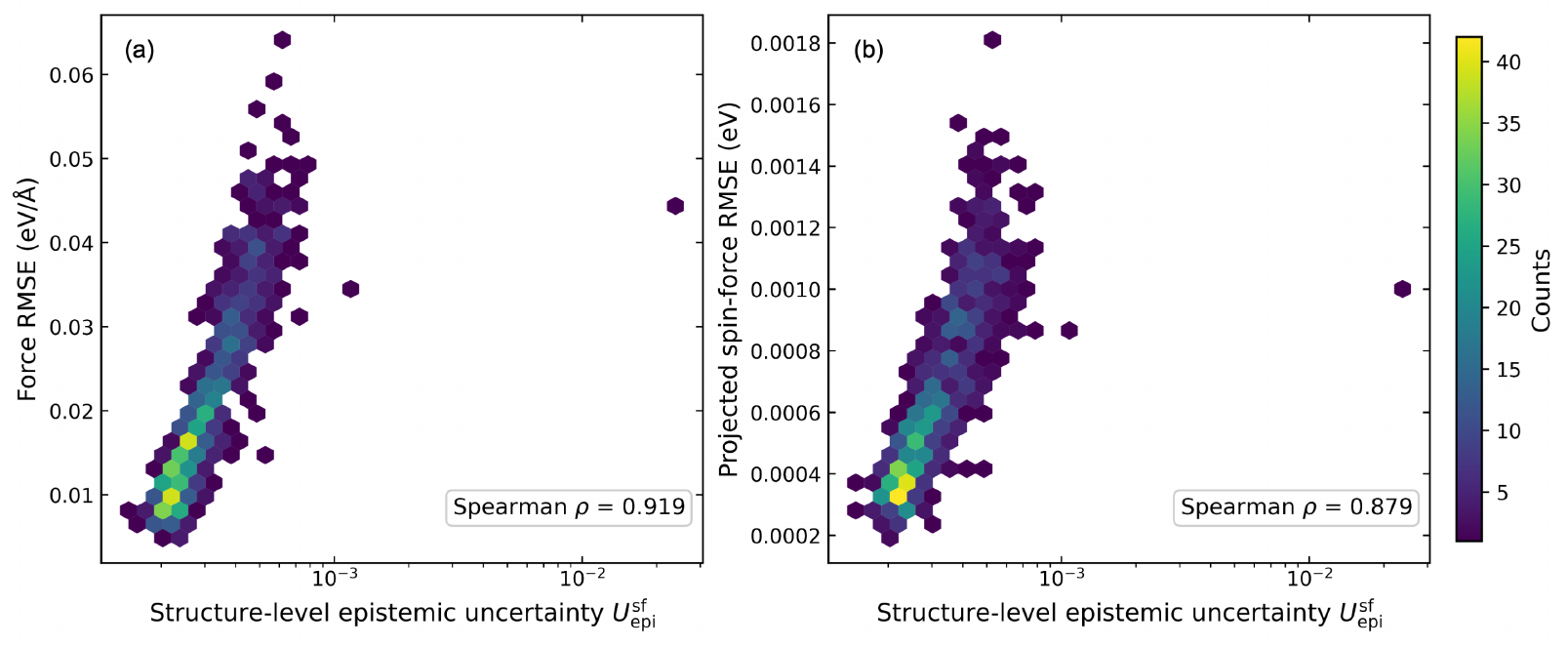}
    \caption{Correlation between the structure-level tangent-plane-consistent epistemic uncertainty $U_{\mathrm{epi}}^{\mathrm{sf}}$ and prediction error in bulk BiFeO$_3$. Panel (a) shows the relation to force RMSE, and panel (b) shows the relation to projected spin-force RMSE. Spearman correlation coefficients are annotated in each panel.}
    \label{fig:bfo-correlation}
\end{figure*}

To assess the usefulness of $U_{\mathrm{epi}}^{\mathrm{sf}}$ as a configuration-selection signal, we compare it with atomic-force and projected-spin-force prediction errors on held-out test configurations. For bulk BiFeO$_3$, Fig.~\ref{fig:bfo-correlation} shows that the tangent-plane-consistent uncertainty score is strongly aligned with both force error and projected spin-force error, with Spearman correlation coefficients of 0.919 and 0.879, respectively. The lower-left to upper-right trend in both panels indicates that configurations assigned larger $U_{\mathrm{epi}}^{\mathrm{sf}}$ generally have larger prediction errors, which is the desired behavior for an active-learning selection score. For monolayer CrTe$_2$, the corresponding correlations are shown in Fig.~S2 of the Supplemental Material\cite{supplemental_material}, with Spearman coefficients of 0.857 for force RMSE and 0.848 for projected spin-force RMSE.

As a control, we also evaluate a force-branch uncertainty score, denoted $U_{\mathrm{epi}}^{f}$. This score follows the original force-target $\mathrm{e}^2\mathrm{IP}$ formulation for nonmagnetic interatomic potentials\cite{wang_equivariant_2026}. We test whether it also ranks magnetic-response errors. Its Spearman correlation with force RMSE is $\rho=0.969$ for BiFeO$_3$ and $\rho=0.930$ for CrTe$_2$. Its correlation with projected spin-force RMSE is lower than that of $U_{\mathrm{epi}}^{\mathrm{sf}}$, with $\rho=0.786$ for BiFeO$_3$ and $\rho=0.772$ for CrTe$_2$. These comparisons, shown in Figs.~S1 and~S2 of the Supplemental Material\cite{supplemental_material}, support the use of $U_{\mathrm{epi}}^{\mathrm{sf}}$ for ranking magnetic-response errors.

\subsection[Active learning performance on bulk BiFeO3]{Active learning performance on bulk BiFeO$_3$}
Bulk BiFeO$_3$ is a room-temperature multiferroic with coupled lattice and G-type antiferromagnetic degrees of freedom\cite{fischer_temperature_1980,wang_epitaxial_2003,yang_deep_2024}, providing a benchmark for testing uncertainty-guided selection in a coupled spin--lattice system. The comparison starts from 1000 shared initial configurations and adds two rounds of 500 configurations. The active-learning dataset uses uncertainty-based selection in each round, whereas the random baseline uses random sampling. The final training set contains 2000 configurations, and the test set contains 878 configurations.

\begin{figure*}
    \centering
    \includegraphics[width=\textwidth]{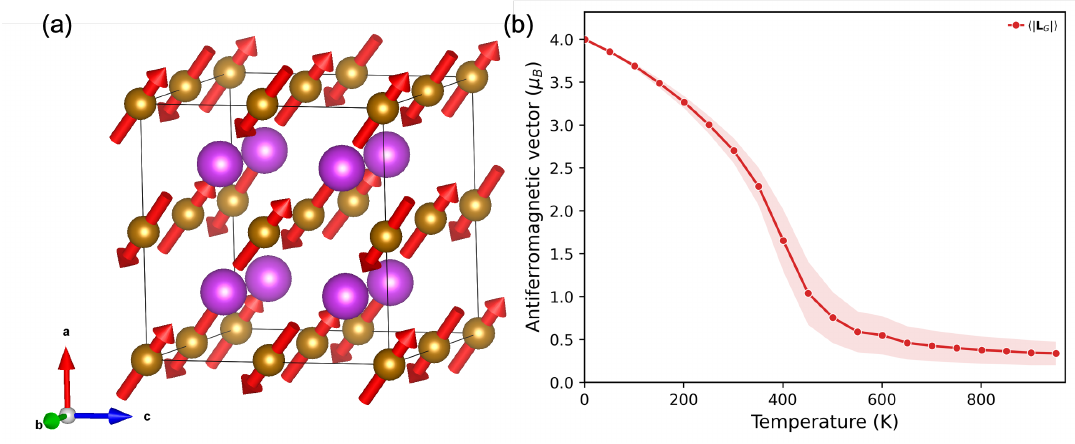}
    \caption{Bulk BiFeO$_3$: spin-lattice structure and magnetic transition. (a) Representative BiFeO$_3$ structure extracted directly from the final spin-lattice-dynamics frame and shown as a pseudocubic $2\times2\times2$ region without symmetry refinement or structural idealization. Fe atoms are shown in brown, Bi atoms in purple, and local spin vectors as red arrows indicating the G-type antiferromagnetic order; the crystallographic axes $a$, $b$, $c$ are shown. (b) Temperature dependence of the G-type antiferromagnetic order parameter $\langle|\mathbf L_G|\rangle$, obtained from spin-lattice-dynamics simulations with the trained potential. The shaded band denotes the standard deviation over the production trajectory at each temperature.}
    \label{fig:bfo-groundphase}
\end{figure*}

The $\mathrm{e}^2\mathrm{SLP}$ active-learning strategy consistently improves all reported error metrics relative to random sampling. The per-atom energy mean absolute error decreases from $0.286$~meV/atom in the random baseline to $0.182$~meV/atom with active learning. The force MAE decreases from $20.0$ to $15.4$~meV/\AA, and the projected spin-force MAE decreases from $0.579$ to $0.438$~meV. These values are summarized in Table~\ref{tab:active-learning-performance}. Figure~\ref{fig:bfo-groundphase} further shows finite-temperature SLD results for a 1280-atom BiFeO$_3$ supercell cooled from 1001 to 1~K. The collapse of the G-type antiferromagnetic order parameter provides a qualitative dynamical consistency check that the trained potential captures the loss of antiferromagnetic order at elevated temperature. Full simulation details are given in the Supplemental Material\cite{supplemental_material}.

\subsection[Active learning performance on monolayer CrTe2]{Active learning performance on monolayer CrTe$_2$}
Monolayer CrTe$_2$ provides a complementary two-dimensional magnetic benchmark with strong spin--lattice coupling\cite{sun_room_2020,zhang_room-temperature_2021,yu_physics-informed_2024}. The comparison starts from a shared initial dataset and uses the same temperature-stratified candidate pool at each expansion round. The methods are compared at training set sizes of 1000, 2000, 2500, and 3000 configurations. The random baseline selects additional configurations randomly and trains the conventional magnetic potential, whereas the committee baseline uses four independently initialized models for ensemble-disagreement selection and trains the same conventional potential. In the $\mathrm{e}^2\mathrm{SLP}$-based branch, configurations are selected using the structure-level $U_{\mathrm{epi}}^{\mathrm{sf}}$ score. The selected data are used to train the evidential $\mathrm{e}^2\mathrm{SLP}$ model, after which a conventional output-head model is initialized from the shared $\mathrm{e}^2\mathrm{SLP}$ parameters and fine-tuned for final accuracy evaluation. The final test set contains 810 configurations.

\begin{figure*}
    \centering
    \includegraphics[width=\textwidth]{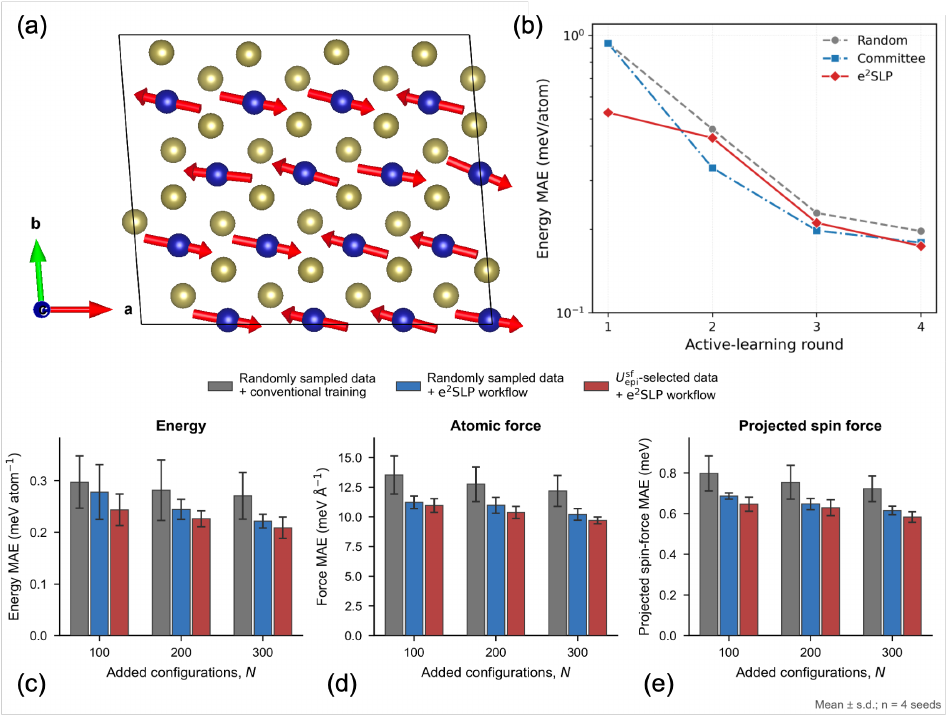}
    \caption{Monolayer CrTe$_2$: structure and active-learning performance. (a) Representative monolayer CrTe$_2$ structure, with Cr atoms (blue), Te atoms (gold), and local spin vectors (red arrows). (b) Per-atom energy MAE versus active-learning round for random sampling, four-model committee disagreement, and the $\mathrm{e}^2\mathrm{SLP}$ active-learning branch. (c)--(e) Ablation on a 500-configuration initial CrTe$_2$ set for energy MAE, atomic-force MAE, and projected-spin-force MAE. In each round, 100 configurations are added and selected either randomly or by $U_{\mathrm{epi}}^{\mathrm{sf}}$. Bars compare random selection with conventional training, random selection with $\mathrm{e}^2\mathrm{SLP}$ training followed by fine-tuning, and $U_{\mathrm{epi}}^{\mathrm{sf}}$-guided selection with $\mathrm{e}^2\mathrm{SLP}$ training followed by fine-tuning. Error bars denote mean $\pm$ s.d. over four seeds.}
    \label{fig:crte2-al}
\end{figure*}

Figure~\ref{fig:crte2-al}(a) shows a representative monolayer CrTe$_2$ structure, and Fig.~\ref{fig:crte2-al}(b) compares the per-atom energy MAE across active-learning rounds. At the final dataset size, the $\mathrm{e}^2\mathrm{SLP}$ active-learning branch outperforms the random baseline on all reported metrics: the energy MAE decreases from $0.197$ to $0.174$~meV/atom, the force MAE from $9.8$ to $8.0$~meV/\AA, and the projected spin-force MAE from $0.540$ to $0.455$~meV. These values are summarized in Table~\ref{tab:active-learning-performance}. At the initial 1000-configuration stage, the $\mathrm{e}^2\mathrm{SLP}$ active-learning branch already gives a lower energy MAE than the conventional models trained on the same initial set ($0.527$ versus $0.937$~meV/atom). With additional configurations, committee selection improves over random sampling and reaches $0.179$~meV/atom at 3000 configurations, while the $\mathrm{e}^2\mathrm{SLP}$ active-learning branch reaches $0.174$~meV/atom. Complete round-by-round results are reported in the Supplemental Material\cite{supplemental_material}.

To further distinguish the contribution of uncertainty-guided selection from that of $\mathrm{e}^2\mathrm{SLP}$ training and fine-tuning, we performed a CrTe$_2$ ablation starting from a 500-configuration initial set. Each round adds 100 configurations selected either randomly or by the structure-level $U_{\mathrm{epi}}^{\mathrm{sf}}$ score. The ablation was evaluated on an independent 500-configuration CrTe$_2$ test set. The comparison includes three settings: random selection with conventional training, random selection with $\mathrm{e}^2\mathrm{SLP}$ training followed by fine-tuning, and $U_{\mathrm{epi}}^{\mathrm{sf}}$-guided selection with $\mathrm{e}^2\mathrm{SLP}$ training followed by fine-tuning. Figure~\ref{fig:crte2-al}(c)--(e) reports the mean and standard deviation over four seeds for energy, force, and projected spin-force MAE.

\begin{table*}
    \centering
    \caption{Performance comparison between random sampling and active learning for the two benchmark magnetic systems, measured by mean absolute error. The lower error in each metric for a given system is highlighted in bold.}
    \label{tab:active-learning-performance}
    \begin{tabular*}{\textwidth}{@{\extracolsep{\fill}}llccc}
        \hline
        System & Sampling strategy & \shortstack{Energy \\ (meV/atom)} & \shortstack{Force \\ (meV/\AA)} & \shortstack{Projected spin force \\ (meV)} \\
        \hline
        BiFeO$_3$ & Random & 0.286 & 20.0 & 0.579 \\
        BiFeO$_3$ & $\mathrm{e}^2\mathrm{SLP}$ active learning & \textbf{0.182} & \textbf{15.4} & \textbf{0.438} \\
        CrTe$_2$ & Random & 0.197 & 9.8 & 0.540 \\
        CrTe$_2$ & $\mathrm{e}^2\mathrm{SLP}$ active learning & \textbf{0.174} & \textbf{8.0} & \textbf{0.455} \\
        \hline
    \end{tabular*}
\end{table*}

\subsection{Discussion}
The improvement in accuracy observed here has two possible sources.
One is configuration selection.
The correlation tests show that $U_{\mathrm{epi}}^{\mathrm{sf}}$ ranks configurations by prediction error in both BiFeO$_3$ and CrTe$_2$.
When this score is used for configuration selection, the final errors decrease relative to random sampling.
In BiFeO$_3$, $U_{\mathrm{epi}}^{\mathrm{sf}}$-based selection lowers the final energy, force, and projected spin-force MAEs relative to random sampling.
In CrTe$_2$, the same trend appears in the full round-by-round comparison.
The smaller four-seed ablation further shows that $U_{\mathrm{epi}}^{\mathrm{sf}}$-guided selection lowers the mean errors relative to random selection with the same $\mathrm{e}^2\mathrm{SLP}$ training and fine-tuning route.

The other possible source is the model and training route.
The original $\mathrm{e}^2\mathrm{IP}$ study showed that the design of the evidential force head and loss can affect force accuracy and data efficiency in interatomic-potential training\cite{wang_equivariant_2026}.
Our CrTe$_2$ controls provide the direct comparison in the present magnetic setting.
With random selection, $\mathrm{e}^2\mathrm{SLP}$ training followed by fine-tuning improves over random selection with non-evidential training.
The lower CrTe$_2$ error at the initial 1000-configuration stage also points to a contribution from the model and fine-tuning protocol, before any difference in newly selected configurations is introduced.

A direct comparison with a three-dimensional spin-force likelihood is not included, since the constrained-moment labels used here define only the transverse spin response.
The present tests are still limited to two benchmark systems, one SLD-based candidate-generation protocol, and one structure-level selection score.
Further work should test other magnetic orders, materials with different spin--lattice coupling mechanisms, alternative selection rules, and longer downstream spin-lattice simulations.

\section{Conclusion}
In this work, we introduce $\mathrm{e}^2\mathrm{SLP}$, an uncertainty-aware method for magnetic spin--lattice potentials inspired by $\mathrm{e}^2\mathrm{IP}$ and developed around a tangent-plane projected-spin-force formulation. By formulating the evidential uncertainty model in the tangent plane of the projected spin-force target, $\mathrm{e}^2\mathrm{SLP}$ provides a structure-level uncertainty score for magnetic-configuration selection. Across bulk BiFeO$_3$ and monolayer CrTe$_2$, $U_{\mathrm{epi}}^{\mathrm{sf}}$ correlates with prediction error and improves active-learning performance relative to random sampling. The CrTe$_2$ controls further show that the accuracy gain comes from both uncertainty-guided selection and the $\mathrm{e}^2\mathrm{SLP}$ training/fine-tuning route. These results support tangent-plane evidential uncertainty as a practical selection signal for data-efficient construction of magnetic machine-learning interatomic potentials.

\begin{acknowledgments}
We acknowledge financial support from the National Key R\&D Program of China (Grant No.~2022YFA1402901), the National Natural Science Foundation of China (NSFC, Grant No.~12188101), the Shanghai Science and Technology Program (No.~23JC1400900), the Guangdong Major Project of Basic and Applied Basic Research (Future Functional Materials under Extreme Conditions, Grant No.~2021B0301030005), the Shanghai Pilot Program for Basic Research at Fudan University (No.~23TQ017), the robotic AI-Scientist platform of the Chinese Academy of Sciences, and the New Cornerstone Science Foundation.
\end{acknowledgments}

\bibliography{refs}

\end{document}

\title{Supplemental Material:\\
Tangent-Plane Evidential Uncertainty in Active Learning for Magnetic Interatomic Potentials}

\author{Yang Cheng}
\affiliation{Key Laboratory of Computational Physical Sciences (Ministry of Education), Institute of Computational Physical Sciences, State Key Laboratory of Surface Physics, and Department of Physics, Fudan University, Shanghai, 200433, China}

\author{Hongyu Yu}
\email[Corresponding author: ]{hongyuyu20@fudan.edu.cn}
\affiliation{Key Laboratory of Computational Physical Sciences (Ministry of Education), Institute of Computational Physical Sciences, State Key Laboratory of Surface Physics, and Department of Physics, Fudan University, Shanghai, 200433, China}

\author{Hongjun Xiang}
\email[Corresponding author: ]{hxiang@fudan.edu.cn}
\affiliation{Key Laboratory of Computational Physical Sciences (Ministry of Education), Institute of Computational Physical Sciences, State Key Laboratory of Surface Physics, and Department of Physics, Fudan University, Shanghai, 200433, China}

\maketitle

%------------------------------------------------------------------
\section{First-principles reference data}
\label{sec:dft}

All reference labels used in this work were generated by spin-polarized noncollinear density-functional theory (DFT) calculations performed with the Vienna \emph{ab initio} Simulation Package (VASP)\cite{kresse_efficiency_1996} using the projector augmented-wave (PAW) method\cite{blochl_projector_1994} and the Perdew--Burke--Ernzerhof (PBE) generalized-gradient functional\cite{perdew_generalized_1996}. Fully noncollinear magnetism with spin--orbit coupling was included throughout to enable the time-reversal-equivariant treatment of magnetic interactions adopted in the SpinGNN++-style backbone\cite{yu_physics-informed_2024}. System-specific parameters are summarized in the following two subsections.

\subsection{Bulk BiFeO$_3$}
\label{sec:dft-bfo}

Bulk BiFeO$_3$ calculations were performed on an 80-atom $2\times2\times2$ supercell of the rhombohedral $R3c$ primitive cell. A plane-wave kinetic-energy cutoff of 520\,eV was used together with a $\Gamma$-centered $3\times3\times3$ Monkhorst--Pack $k$-mesh. Brillouin-zone integration used Gaussian smearing with a width of 0.05\,eV. An effective Hubbard correction of $U_\mathrm{eff}=3.8$\,eV was applied to the Fe-$3d$ shell.

The projected spin-force labels $\mathbf f_{s,\perp}$ used to supervise the magnetic branch were obtained from constrained noncollinear DFT, following the constrained-functional formulation of Ma and Dudarev\cite{ma_constrained_2015,zhang_fully_2026}. The spin-force target corresponds to the derivative of the energy with respect to the dimensionless local spin-direction unit vector $\hat{\mathbf s}_i$, so it has units of energy. In these direction-only constrained calculations, the target local-moment directions were prescribed, while the magnetic-moment magnitudes were allowed to relax self-consistently. The constraining penalty acts in the subspace transverse to the local magnetic moment, so the residual transverse constraining field served as the magnetic-supervision signal used in training. We used a transverse penalty strength of $\lambda=10$. In our convention, the per-site DFT target is defined from this transverse constraining field as $\mathbf f_{s,\perp,i}^{\mathrm{DFT}}=-|\mathbf M_i|\,\mathbf B^{\mathrm{constr}}_{\perp,i}$, where $|\mathbf M_i|$ is the converged local magnetic-moment magnitude from the constrained DFT calculation and converts the transverse constraining field to the energy scale used for the spin-direction derivative.

\subsection{Monolayer CrTe$_2$}
\label{sec:dft-crte2}

Monolayer CrTe$_2$ calculations were performed on a 48-atom Cr$_{16}$Te$_{32}$ supercell. A plane-wave kinetic-energy cutoff of 400\,eV was used together with a $6\times6\times1$ Monkhorst--Pack $k$-mesh and an out-of-plane vacuum spacing of 20.17\,\AA{}.
On-site correlations on the Cr-$3d$ shell were treated using the rotationally invariant DFT+U formalism in the Liechtenstein form\cite{liechtenstein_density-functional_1995}, with $U=3$\,eV and $J=0.6$\,eV.

Following the same direction-only constrained DFT procedure as for BiFeO$_3$, the projected spin-force labels $\mathbf f_{s,\perp}$ used to supervise the magnetic branch were obtained from constrained noncollinear DFT\cite{ma_constrained_2015,zhang_fully_2026}. The target local-moment directions were prescribed, while the magnetic-moment magnitudes were allowed to relax self-consistently. We used a transverse penalty strength of $\lambda=5$. The residual transverse constraining field served as the magnetic-supervision signal used in training, and the converged local moment magnitude was used in the same spin-force convention as for BiFeO$_3$.

%------------------------------------------------------------------
\section{Initial datasets and active-learning protocol}
\label{sec:dataset}

\subsection{Bulk BiFeO$_3$}

The initial training pool for bulk BiFeO$_3$ contained 1000 noncollinear DFT-labeled configurations. Atomic configurations were sampled from low-accuracy \emph{ab initio} molecular-dynamics (AIMD) trajectories of the 80-atom $R3c$ BiFeO$_3$ supercell described in Sec.~\ref{sec:dft-bfo}, run at temperatures spanning 200--1000\,K. For each sampled configuration the Fe magnetic moments were re-initialized with random noncollinear orientations before performing a single-point constrained noncollinear DFT calculation with the parameters of Sec.~\ref{sec:dft-bfo}. This procedure decorrelates the spin and lattice degrees of freedom in the initial pool and exposes the model to non-equilibrium magnetic configurations from the very first training round.

The training set was then built through two rounds of active learning, yielding a final training set of 2000 configurations. The held-out test set contained 878 configurations.

\subsection{Monolayer CrTe$_2$}

The initial training pool for monolayer CrTe$_2$ contained 1000 noncollinear DFT-labeled configurations. Atomic configurations were sampled from low-accuracy \emph{ab initio} molecular-dynamics (AIMD) trajectories of the 48-atom Cr$_{16}$Te$_{32}$ supercell described in Sec.~\ref{sec:dft-crte2}, run at temperatures spanning 1--300\,K. For each sampled configuration the Cr magnetic moments were re-initialized with random noncollinear orientations before performing a single-point constrained noncollinear DFT calculation with the parameters of Sec.~\ref{sec:dft-crte2}. This procedure decorrelates the spin and lattice degrees of freedom in the initial pool and exposes the model to non-equilibrium magnetic configurations from the very first training round.

The training set was then built through four active-learning stages with training set sizes of 1000, 2000, 2500, and 3000 configurations. The held-out test set contained 810 configurations. The same 810-configuration test set was used for all round-by-round CrTe$_2$ accuracy comparisons.

\subsection{Candidate-generation protocol}
\label{sec:candidate-generation}

At each active-learning round, a common temperature-stratified candidate pool was generated by spin-lattice dynamics (SLD) simulations\cite{tranchida_massively_2018,thompson_lammps_2022,garcia-palacios_langevin-dynamics_1998} using the latest $\mathrm{e}^2\mathrm{SLP}$ model. The candidate-generation model was trained on the common initial dataset for the first expansion and on the $\mathrm{e}^2\mathrm{SLP}$-selected dataset thereafter. Random, committee-based (for CrTe$_2$), and $\mathrm{e}^2\mathrm{SLP}$-based acquisition were then applied to this identical pool using the same per-temperature selection budget. Candidate generation was thus held fixed across acquisition strategies, so the comparison isolates the acquisition criterion rather than differences between strategy-specific candidate pools.

For bulk BiFeO$_3$, the common pool was stratified by temperature, with eight windows from 200--300 to 900--1000\,K. Each temperature window contributed approximately 160 candidate configurations. In each round, 500 configurations were selected for DFT labeling, with 62 or 63 configurations chosen from each temperature window. The random and $\mathrm{e}^2\mathrm{SLP}$ strategies used the same temperature-stratified candidate pool and the same per-window selection counts. They differed only in the acquisition score: random sampling selected configurations uniformly within each window, whereas $\mathrm{e}^2\mathrm{SLP}$ selection ranked candidates by $U_{\mathrm{epi}}^{\mathrm{sf}}$ defined in Eqs.~(16)--(17) of the main text.

For monolayer CrTe$_2$, active-learning candidate generation used ten temperature windows, 1--31, 31--61, ..., 271--301\,K. At each expansion round, each window contributed 300 candidate configurations to the common pool. Starting from the common 1000-configuration initial dataset described above, the first active-learning expansion selected 100 configurations from each window, adding 1000 configurations and producing the 2000-configuration dataset. The next two expansions selected 50 configurations from each window, adding 500 configurations per round and producing the 2500- and 3000-configuration datasets. The random, committee, and $\mathrm{e}^2\mathrm{SLP}$ strategies used the same temperature-stratified candidate pools and differed only in the acquisition score.

\subsection{Finite-temperature BiFeO$_3$ spin-lattice simulation}
\label{sec:bfo-sld-phase}

The magnetic transition shown in Fig.~3 of the main text was obtained from the production LAMMPS input used for the BiFeO$_3$ phase simulation\cite{tranchida_massively_2018,thompson_lammps_2022}. The simulation used the trained BFO magnetic potential and a 1280-atom periodic supercell generated by replicating the 80-atom $R3c$ BiFeO$_3$ cell by $2\times4\times2$. This supercell contains $N=256$ magnetic Fe sites. Dynamics were propagated using spin-lattice NPT dynamics in LAMMPS, with a time step of 0.0005\,ps, lattice thermostat damping time $100\Delta t$, pressure damping time $1000\Delta t$, spin damping parameter 0.4, and zero external pressure.

The system was initialized at 1001\,K and then cooled in 50\,K intervals down to 1\,K. At each temperature, the protocol included an 8\,ps cooling segment from the previous temperature, a 4\,ps equilibration segment at the target temperature, and a 30\,ps production segment used for collecting magnetic observables. Nonmagnetic Bi and O atoms were assigned zero spin force during the SLD run.

The structural snapshot shown in Fig.~3a of the main text was extracted from the final SLD frame. For visualization, the final-frame atomic coordinates were periodically replicated and cropped to a pseudocubic $2\times2\times2$ region; no symmetry refinement or idealized $R3c$ coordinate reconstruction was applied.

The G-type antiferromagnetic order parameter was computed from the two Fe magnetic sublattices defined in the 80-atom parent cell and then replicated with the simulation cell. Denoting the sublattice-averaged spin vectors by $\langle\mathbf S\rangle_{g+}$ and $\langle\mathbf S\rangle_{g-}$, we used
\begin{equation}
\mathbf L_G=\frac{\langle\mathbf S\rangle_{g+}-\langle\mathbf S\rangle_{g-}}{2},
\end{equation}
We monitored the instantaneous scalar order parameter $|\mathbf L_G|$ over the production trajectory at each temperature and reported the trajectory average $\langle|\mathbf L_G|\rangle$ together with its standard deviation in Fig.~3b of the main text.

%------------------------------------------------------------------
\section{$\mathrm{e}^2\mathrm{SLP}$ model details}
\label{sec:model}

\subsection{Backbone hyperparameters}
\label{sec:backbone-hp}

The SOC-aware backbone follows the SpinGNN++ decomposition\cite{yu_physics-informed_2024} and is implemented through a Spin-Allegro-style local-environment network\cite{musaelian_learning_2023}. Both the magnetic-symmetry-preserving channel ($E^{\mathrm{MSENN}}$) and the time-reversal-equivariant channel ($E^{\mathrm{TENN}}$) share the atomic graph but use distinct learnable parameters. Backbone hyperparameters are shared between bulk BiFeO$_3$ and monolayer CrTe$_2$ except for the depth of the latent MLP stack, as detailed below. We adopt a local cutoff radius of $r_c=7.5$\,\AA{} and an O(3)-equivariant local-environment representation with time-reversal parity channels, with the maximum tensor rank truncated at $\ell_\mathrm{max}=2$ for both the equivariant features and the spin--orbit-coupled channel. The network comprises two Allegro tensor-product layers in the rotation-only branch and two in the spin--orbit branch, with an environment-embedding multiplicity of $24$ and four channels for the $\mathrm{e}^2\mathrm{SLP}$ latent and tensor-square features. Edge distances are encoded by eight trainable Bessel functions multiplied by a polynomial cutoff envelope of order $p=48$. The two-body latent MLP has hidden dimensions $[64,128,256]$, with subsequent latent MLPs $[256,256]$ for BiFeO$_3$ and $[256,256,256]$ for CrTe$_2$; the edge-energy MLPs use hidden dimensions $[128,128,64]$ and the node MLPs hidden dimension $128$; SiLU activations are applied throughout the radial and latent MLPs, while the equivariant node MLPs use linear activations.

\subsection{Evidential heads}
\label{sec:evidential-heads}

Following the equivariant evidential construction of Wang \emph{et al.}\cite{wang_equivariant_2026} and the evidential-regression framework of Amini \emph{et al.}\cite{amini_deep_nodate,meinert_multivariate_2022}, the predictive mean for the projected-spin-force branch is not produced by an independent vector head. Instead, it is obtained from the projected energy gradient with respect to the local spin-direction unit vector, $\boldsymbol\gamma_i=-\mathbf P_i\,\partial E/\partial \hat{\mathbf s}_i$, as described in the main text. The evidential heads therefore parameterize only the evidence parameters $\kappa_i$ and $\nu_i$ and the symmetric matrix $\mathbf S_i$ for the projected-spin-force branch. The symmetric matrix $\mathbf S_i$ is mapped to the canonical uncertainty-shape matrix through the matrix exponential,
\begin{equation}
\boldsymbol\Sigma_{0,i}=\exp(\mathbf S_i),
\end{equation}
guaranteeing strict positive-definiteness while preserving rotation-equivariance. For the projected spin-force branch, $\boldsymbol\Sigma_{0,i}$ is then projected into the local tangent plane,
\begin{equation}
\boldsymbol\Sigma_{0,i,\perp}=\mathbf P_i\,\boldsymbol\Sigma_{0,i}\,\mathbf P_i,
\qquad
\mathbf P_i=\mathbf I-\hat{\mathbf s}_i\hat{\mathbf s}_i^{\top},
\end{equation}
and the multivariate Student-$t$ likelihood is evaluated in two-dimensional tangent coordinates with $d=2$, as detailed in the main text.

\subsection{Training procedure}
\label{sec:training}

The total loss combines the magnetic negative log-likelihood term given in the main text with standard energy and force losses,
\begin{equation}
\mathcal L_\mathrm{tot}
= \lambda_E\,\mathcal L_E
+ \lambda_F\,\mathcal L_F
+ \lambda_\mathrm{sf}\,\mathcal L_\mathrm{NLL}^\mathrm{sf}
+ \lambda_\mathrm{ev}\,\mathcal L_\mathrm{ev},
\end{equation}
where $\mathcal L_E$ takes a per-atom mean-squared-error form. For the projected-spin-force evidential branch, we use the error-weighted evidence penalty\cite{amini_deep_nodate,wang_equivariant_2026}
\begin{equation}
\mathcal L_\mathrm{ev}
=\frac{1}{N_\mathrm{mag}}\sum_{i\in \mathcal C_\mathrm{mag}}
(\nu_i+\kappa_i)
\left\|
\mathbf y_{i,2d}-\boldsymbol\gamma_{i,2d}
\right\|_2,
\label{eq:evidence-regularizer}
\end{equation}
where $\mathcal C_\mathrm{mag}$ denotes the magnetic sites in the training structure and $N_\mathrm{mag}=|\mathcal C_\mathrm{mag}|$. Thus, large evidence is penalized only when the tangent-plane projected-spin-force prediction error is large.

For both bulk BiFeO$_3$ and monolayer CrTe$_2$ we use loss weights $\lambda_E:\lambda_F:\lambda_\mathrm{sf}:\lambda_\mathrm{ev}=40:10:0.002:0.001$ and optimize $\mathcal L_\mathrm{tot}$ with Adam (AMSGrad enabled, $\beta_1=0.9$, $\beta_2=0.999$, $\varepsilon=10^{-8}$, zero weight decay), an initial learning rate of $2\times10^{-3}$, and a batch size of $1$.

%------------------------------------------------------------------
\section{Additional uncertainty--error correlations}
\label{sec:additional-correlations}

For the force-only evidential ablation, we implemented in our code the force-target $\mathrm{e}^2\mathrm{IP}$ formulation introduced by Wang \emph{et al.}\cite{wang_equivariant_2026}. The evidential NLL and evidence regularizer were applied to the atomic-force target with $d=3$ and a full $3\times3$ SPD covariance tensor, while the projected-spin-force evidential branch was removed. The force-branch epistemic score was converted to a structure-level scalar by the trace average
\begin{equation}
U_{\mathrm{epi}}^{f}(\mathcal C)
=\frac{1}{|\mathcal C|}\sum_{i\in\mathcal C}
\mathrm{tr}\!\left(\mathbf U_{\mathrm{epi},i}^{f}\right).
\label{eq:force-branch-epistemic-score}
\end{equation}
The projected-spin-force errors reported for this ablation are therefore used only to test whether a force-trained uncertainty signal also ranks magnetic-response errors.

\begin{figure}
    \centering
    \includegraphics[width=\columnwidth]{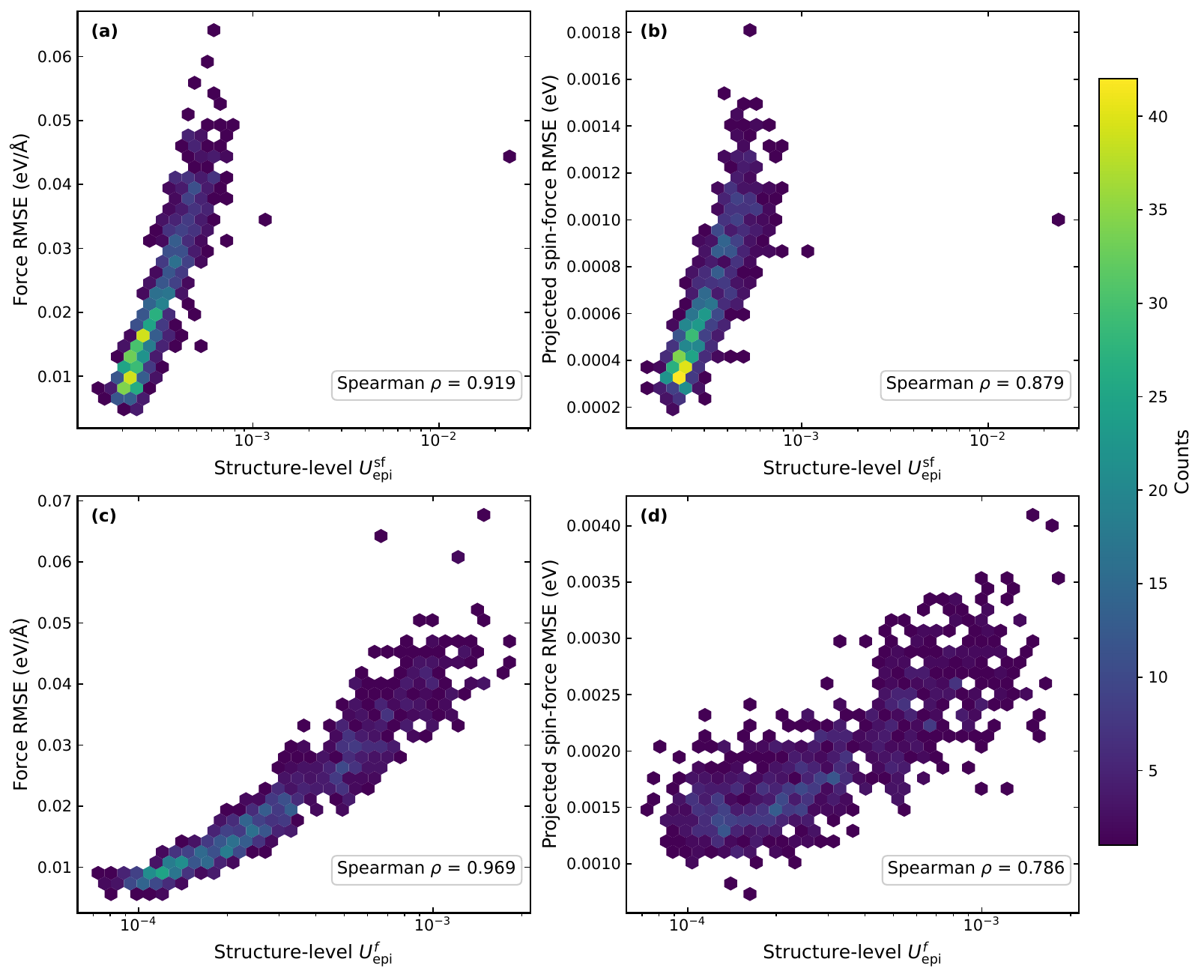}
    \caption{Bulk BiFeO$_3$: structure-level epistemic uncertainty versus prediction error for the projected-spin-force indicator $U_{\mathrm{epi}}^{\mathrm{sf}}$ (top row) and the force-branch indicator $U_{\mathrm{epi}}^{f}$ (bottom row). Columns: (a, c) force RMSE; (b, d) projected spin-force RMSE. Both uncertainty indicators are covariance-trace scores with the squared units of their respective vector targets and are used as relative ranking indicators. Spearman $\rho$ is annotated in each panel; the color scale is shared across all four panels.}
    \label{fig:bfo-uepi-comparison-si}
\end{figure}

\begin{figure}
    \centering
    \includegraphics[width=\columnwidth]{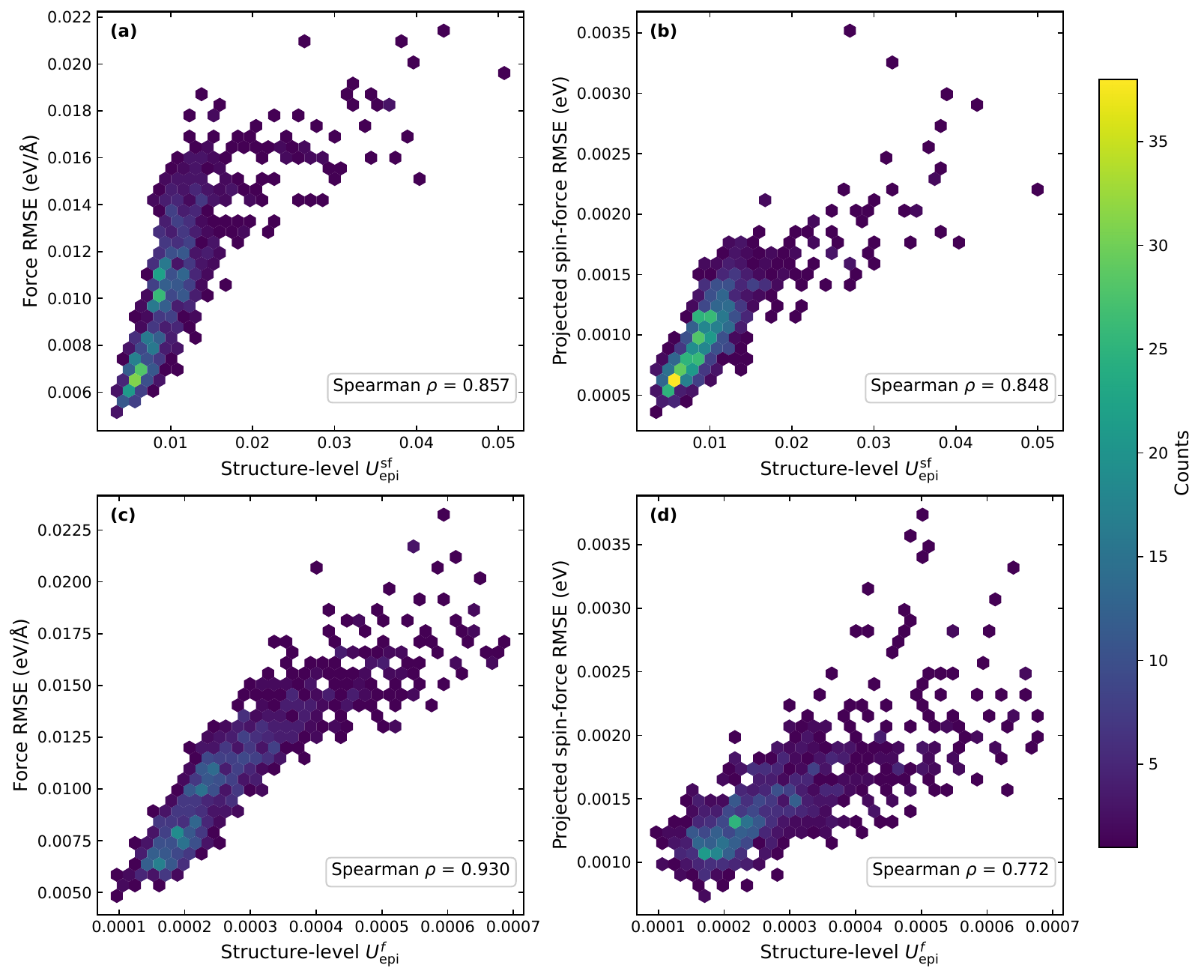}
    \caption{Monolayer CrTe$_2$: structure-level epistemic uncertainty versus prediction error for the projected-spin-force indicator $U_{\mathrm{epi}}^{\mathrm{sf}}$ (top row) and the force-branch indicator $U_{\mathrm{epi}}^{f}$ (bottom row). Columns: (a, c) force RMSE; (b, d) projected spin-force RMSE. Both uncertainty indicators are covariance-trace scores with the squared units of their respective vector targets and are used as relative ranking indicators. Spearman $\rho$ is annotated in each panel; the color scale is shared across all four panels.}
    \label{fig:crte2-uepi-comparison-si}
\end{figure}

Figures~\ref{fig:bfo-uepi-comparison-si} and~\ref{fig:crte2-uepi-comparison-si} present, for each benchmark system, the projected-spin-force epistemic indicator $U_{\mathrm{epi}}^{\mathrm{sf}}$ and the force-branch epistemic indicator $U_{\mathrm{epi}}^{f}$ each plotted against the per-structure force RMSE and projected spin-force RMSE. The top row of each figure is the $U_{\mathrm{epi}}^{\mathrm{sf}}$ counterpart of the main-text correlation analysis: for monolayer CrTe$_2$, panels (a) and (b) of Fig.~\ref{fig:crte2-uepi-comparison-si} reproduce the same uncertainty--error alignment seen in bulk BiFeO$_3$, with Spearman $\rho=0.857$ against force RMSE and $\rho=0.848$ against projected spin-force RMSE.
The bottom row of each figure compares the force-only evidential ablation, trained without the projected-spin-force evidential branch. The force-only $U_{\mathrm{epi}}^{f}$ remains strongly correlated with force RMSE (BFO $\rho=0.969$; CrTe$_2$ $\rho=0.930$) but its alignment with projected spin-force RMSE drops markedly relative to $U_{\mathrm{epi}}^{\mathrm{sf}}$ (BFO $0.879\!\rightarrow\!0.786$; CrTe$_2$ $0.848\!\rightarrow\!0.772$). This comparison supports the use of the tangent-plane projected-spin-force uncertainty indicator $U_{\mathrm{epi}}^{\mathrm{sf}}$ for ranking magnetic-response errors in active learning.

\section{\texorpdfstring{C\lowercase{r}T\lowercase{e}$_2$}{CrTe2} active-learning controls and training-route ablation}
\label{sec:crte2-round-controls}

Table~\ref{tab:crte2-round-controls} summarizes the monolayer CrTe$_2$ round-by-round comparison used to interpret Fig.~4(b) of the main text. The four rounds correspond to training set sizes of 1000, 2000, 2500, and 3000 configurations. Random denotes random configuration selection followed by training of the conventional magnetic potential. Committee denotes selection by ensemble disagreement from four independently initialized models, followed by training of the same conventional potential. In the $\mathrm{e}^2\mathrm{SLP}$-based branch, configurations are selected using the structure-level $U_{\mathrm{epi}}^{\mathrm{sf}}$ score. The selected data are used to train the evidential $\mathrm{e}^2\mathrm{SLP}$ model, after which a conventional output-head model is initialized from the shared $\mathrm{e}^2\mathrm{SLP}$ parameters and fine-tuned for final accuracy evaluation.

At the final 3000-configuration dataset size, the $\mathrm{e}^2\mathrm{SLP}$ active-learning branch gives errors close to the committee baseline and lower than random sampling across energy, force, and projected spin force. Because this round-by-round comparison uses a single training run for each branch, the small difference between the $\mathrm{e}^2\mathrm{SLP}$ and committee energy errors is treated as a numerical comparison rather than a statistical ranking.

\begin{table*}
    \centering
    \caption{Round-by-round monolayer CrTe$_2$ active-learning controls. Dataset size denotes the number of training configurations. Energy MAE is reported in meV/atom, force MAE in meV/\AA, and projected spin-force MAE in meV.}
    \label{tab:crte2-round-controls}
    \begin{tabular*}{\textwidth}{@{\extracolsep{\fill}}llcccc}
        \hline
        Metric & Strategy & 1000 & 2000 & 2500 & 3000 \\
        \hline
        Energy & Random & 0.937 & 0.459 & 0.229 & 0.197 \\
        Energy & Committee & 0.937 & 0.333 & 0.198 & 0.179 \\
        Energy & $\mathrm{e}^2\mathrm{SLP}$ & 0.527 & 0.427 & 0.211 & 0.174 \\
        \hline
        Force & Random & 17.8 & 14.8 & 11.3 & 9.8 \\
        Force & Committee & 17.8 & 12.0 & 9.8 & 9.0 \\
        Force & $\mathrm{e}^2\mathrm{SLP}$ & 14.0 & 10.7 & 8.6 & 8.0 \\
        \hline
        Projected spin force & Random & 0.845 & 0.820 & 0.615 & 0.540 \\
        Projected spin force & Committee & 0.845 & 0.650 & 0.532 & 0.502 \\
        Projected spin force & $\mathrm{e}^2\mathrm{SLP}$ & 0.730 & 0.625 & 0.487 & 0.455 \\
        \hline
    \end{tabular*}
\end{table*}

The ablation in Fig.~4(c)--(e) starts from a 500-configuration initial CrTe$_2$ set. Each round adds 100 configurations selected either randomly or by the structure-level $U_{\mathrm{epi}}^{\mathrm{sf}}$ score. The ablation was evaluated on an independent 500-configuration CrTe$_2$ test set. The comparison includes random selection with conventional training, random selection with $\mathrm{e}^2\mathrm{SLP}$ training followed by fine-tuning, and $U_{\mathrm{epi}}^{\mathrm{sf}}$-guided selection with the same $\mathrm{e}^2\mathrm{SLP}$ training and fine-tuning route. Table~\ref{tab:crte2-workflow-ablation} reports the mean and standard deviation over four seeds.

\begin{table*}
    \centering
    \caption{CrTe$_2$ selection and training-route ablation on a 500-configuration initial set, evaluated on an independent 500-configuration test set. The columns denote the number of added configurations. Values are mean $\pm$ s.d. over four seeds. Energy MAE is reported in meV/atom, force MAE in meV/\AA, and projected spin-force MAE in meV.}
    \label{tab:crte2-workflow-ablation}
    \footnotesize
    \begin{tabular*}{\textwidth}{@{\extracolsep{\fill}}lp{0.32\textwidth}ccc}
        \hline
        Metric & Strategy & 100 & 200 & 300 \\
        \hline
        Energy & Random selection + conventional training & $0.297\pm0.051$ & $0.281\pm0.059$ & $0.271\pm0.045$ \\
        Energy & Random selection + $\mathrm{e}^2\mathrm{SLP}$ training/fine-tuning & $0.278\pm0.053$ & $0.244\pm0.020$ & $0.222\pm0.013$ \\
        Energy & $U_{\mathrm{epi}}^{\mathrm{sf}}$-guided selection + $\mathrm{e}^2\mathrm{SLP}$ training/fine-tuning & $0.244\pm0.031$ & $0.226\pm0.015$ & $0.209\pm0.021$ \\
        \hline
        Force & Random selection + conventional training & $13.5\pm1.6$ & $12.8\pm1.5$ & $12.2\pm1.3$ \\
        Force & Random selection + $\mathrm{e}^2\mathrm{SLP}$ training/fine-tuning & $11.2\pm0.5$ & $11.0\pm0.7$ & $10.2\pm0.5$ \\
        Force & $U_{\mathrm{epi}}^{\mathrm{sf}}$-guided selection + $\mathrm{e}^2\mathrm{SLP}$ training/fine-tuning & $11.0\pm0.6$ & $10.4\pm0.5$ & $9.7\pm0.3$ \\
        \hline
        Projected spin force & Random selection + conventional training & $0.798\pm0.086$ & $0.754\pm0.083$ & $0.722\pm0.063$ \\
        Projected spin force & Random selection + $\mathrm{e}^2\mathrm{SLP}$ training/fine-tuning & $0.686\pm0.015$ & $0.647\pm0.027$ & $0.615\pm0.021$ \\
        Projected spin force & $U_{\mathrm{epi}}^{\mathrm{sf}}$-guided selection + $\mathrm{e}^2\mathrm{SLP}$ training/fine-tuning & $0.647\pm0.035$ & $0.629\pm0.039$ & $0.583\pm0.026$ \\
        \hline
    \end{tabular*}
\end{table*}

\clearpage
\bibliography{refs}